\documentstyle {article}
\begin{document}
\thispagestyle{plain}
\def\a{\alpha}
\def\b{\beta}
\def\g{\gamma}
\def\d{\delta}
\def\e{\epsilon}
\def\k{\kappa}
\def\l{\lambda}
\def\x{\xi}
\def\f{\phi}
\def\j{\psi}
\def\z{\zeta}
\def\p{\partial}
\def\o{\omega}
\def\O{\Omega}
\def\Om0{\Omega^0}
\def\O1{\Omega^1}
\def\cA{{\cal A}}
\def\cB{{\cal B}}
\def\tcA{\tilde{\cal A}}
\def\tcB{\tilde{\cal B}}
\def\tf{\tilde{f}}
\def\tg{\tilde{g}}
\def\th{\tilde{h}}
\def\tu{\tilde{u}}
\def\tv{\tilde{v}}
\def\tL{\tilde L}
\def\rarr{\rightarrow}
\def\mp{\mapsto}
\def\hL{\hat{L}}
\def\hu{\hat{u}}
\def\res{{\rm res}}
\def\tr{{\rm tr}\:}
\def\cF{{\cal F}}
\def\et{\eta}
\def\cV{{\cal V}}
\def\>{\rangle}
\def\<{\langle}
\def\Cal{\cal}
\def\({\left(}
\def\){\right)}
\def\[{\left[}
\def\]{\right]}


\vspace{.2in} \large \centerline{{\Large{\bf On the
Treves Criterion for the Boussinesq}}}
\centerline{{\Large{\bf and other GD Hierarchies.}}}

\vspace{.2in}
\centerline{{\bf L. A. Dickey}}

\vspace{.1in}
\centerline{University of Oklahoma}

\vspace {.3in}
\begin{abstract}This is an addition to our paper ({\it Lett.Math.Phys} 65,
2003,187-197) where an analogue of the Treves criterion for the first integrals of the KdV
hierarchy was suggested for the Boussinesq hierarchy and its necessity was proven.
In the present paper, it is proven that there exists the second, ``conjugated'',
group of tests. Besides, the relationship between our method and the B\"acklund
transformation method recently suggested by Morosi and Pizzocchero is discussed.
\end{abstract}

\vspace{.2in} {\bf 1.} This is an addition to our paper [1]. The
Boussinesq hierarchy was discussed there: the set of equations $\p
L/\p t_i=[L^{(i/3)}_+,L]$ where $L=\p^3+u_1\p+u_0$, and $\p=\p/\p
x$. As it is well-known, the first integrals of these equations
are $\res_\p L^{m/3}$ ($\res_\p$ symbolizes the coefficient in
$\p^{-1}$) modulo derivatives of any differential polynomials of
$u_1$ and $u_0$.
The following proposition was proven in [1] ($\res_x$ symbolizes
the coefficient in the term with $x^{-1}$ in a Laurent
series)\footnote{There are misprints in [1]: the limits of
summation in the first sums in both (0.4) and (0.5) are 1 and $\infty$.}:\\

{\bf Proposition.} The differential polynomials $P[u_1,u_0]=\res_\p
L^{m/3}$ satisfy the following two criteria:

(i) $$\res_xP[\tu_1(x),\tu_0(x)]=0\eqno{(1)}$$ when the series
$$\tu_1(x)=-3x^{-2}+\sum_1^\infty u_{1i}(x^i/i!),~\tu_0(x)=3x^{-
3}+\sum_0^\infty u_{0i}(x^i/i!) ~\eqno{(2)}$$ with $u_{01}=u_{12}/2$ are
substituted for variables $u_1$ and $u_0$.

(ii) The same equality {\rm (1)} is true when the series are
$$\tu_1(x)=-6x^{-2}+\sum_1^\infty u_{1i}(x^i/i!),~\tu_0(x)=12x^{-
3}+\sum_0^\infty u_{0i}(x^i/i!),\eqno{(3)}$$ with
$$u_{01}=u_{12}~{\rm and}~u_{02}=2u_{13}/3.$$

This proposition is an analogue of the necessity part of the Treves
theorem for the KdV hierarchy where $L$ is an operator of the second
order, see [2].

We are going to show that besides this group of two tests there
is the second, ``conjugated'', group of two tests (i$^*$) and (ii$^*$)
of which the first one coincides with (i) but the second is different:

(ii$^*$) Eq. (1) is true if $\tu_1(x)$ and $\tu_2(x)$ are replaced by
$$\tu_1^*(x)=-6x^{-2}+\sum_1^\infty u_{1i}
^*(x^i/i!),~\tu_0^*(x)=\sum_0^\infty u_{0i}^*(x^i/i!),\eqno{(4)}$$ with
$u_{01}^*=0$ and $u_{02}^*=u_{13}^*/3.$

Let us prove this. It is shown in [1] that if the series (2) or (3) are
substituted for $u_1$ and $u_0$ in the operator $L$ then the obtained
operator $\tL=\p^3+\tu_1(x)\p+\tu_0(x)$ can be represented in a ``dressing''
form $\tL=w\p^3w^{-1}$ where $w=\sum_0^\infty w_i\p^i$, $w_0=1$ and all
$w_i$ are Laurent series in $x$. The equation (1) follows from this
fact. Let us go to the conjugated operators: $\tL^*=-(w^*)^{-1}\p^3w^*$.
The operator $-\tL^*$ is $$-\tL^*=\p^3+u_1^*\p+u_0^*=(w^*)^{-1}\p^3w^*,
~{\rm where}~u_1^*=u_1, ~u_0^*=-u_0+u_1'.$$ It is represented in the same
dressing form where the dressing operator has the Laurent series
coefficients. Therefore, the equality (1) is true when $\tu_1^*=\tu_1$
and $\tu_0^*=-\tu_0+\tu_1'$ are substituted for $\tu_1$ and $\tu_0$:
$\res_xP[\tu_1^*(x),\tu_0^*(x)]=0$ or $$\res_xP[\tu_1(x),-\tu_0(x)+\tu_1'(x)]
=0. \eqno{(5)}$$ This gives us the second, ``conjugated'', group of
tests.

We find for the test (i): $$\tu_1^*=\tu_1=-3x^{-2}+\sum_1^\infty u_{1i}(x^i/i!)$$
and $$\tu_0^*=-\tu_0+\tu_1'=-3x^{-
3}-\sum_0^\infty u_{0i}x^i/i!+6x^{-3}+\sum_0^\infty u_{1,i+1}x^i/i!$$$$
=3x^{-3}+\sum_0^\infty u_{0i}^*x^i/i!,~u_{0i}^*=-u_{0i}+u_{1,i+1}.$$
In particular, $$u_{01}^*=-u_{01}+u_{12}=-
u_{12}/2+u_{12}=u_{12}/2=u_{12}^*/2.$$ We see that this test is the
same as (i).

Now, for the test (ii): $$\tu_1^*=\tu_1=-6x^{-2}+\sum_1^\infty u_{1i}(x^i/i!)$$
and $$\tu_0^*=-\tu_0+\tu_1'
=-12x^{-3}-\sum_0^\infty u_{0i}(x^i/i!)+12x^{-3}+\sum_0^\infty
u_{1,i+1}x^i/i!$$$$=\sum_0^\infty u_{0i}^*x^i/i!,~u_{0i}^*=-u_{0i}+u_{1,i+1}.$$
In particular, $$u_{01}^*=-u_{01}+u_{12}=0,$$$$u_{0,2}^*=-u_{0,2}+u_{13}=
-2u_{13}/3+u_{13}=u_{13}/3=u_{13}^*/3.$$ We obtained the test (4).

There was a conjecture in [1] that if a differential polynomial
$P[u_1,u_0]$ satisfies conditions (i) and (ii) then it is a linear
combination of $\res_\p L^{m/3}$. It is not clear: maybe all three conditions,
(i), (ii) and (ii$^*$), should be checked for this, maybe not, and it
suffices to check one group of tests, either the first or the conjugated.
We know nothing about the independence of these tests.\\

{\bf 2.} There was an explanation in [1] how the singular parts of the
series (2) and (3) were guessed. More than that, a recipe was given how to find
these singular parts if the operator $L
=\p^n+u_{n-2}\p^{n-2}+...+u_0$
is of an arbitrary order $n$. Namely, the singular part of the operator
$\tL$ (i.e., only the singular terms of the Laurent series $\tu_i$ are
preserved) can be found from the formula
$$\tL_N=\(\p-{1\over x}\)^N\p^n\(\p-{1\over x}\)^{-N}$$$$=
\(\p-{1\over x}\)^N\(\p+{n-1\over x}\)\(\p-{1\over x}\)^{n-1-N},
~0\leq N<n.\eqno{(6)}$$ When $N=0$, we obtain $\tL_0=\p^n$, an operator
without singularities. This operator does not correspond to any test.

{\it Examples.} Let $n=2$ and $N=1$. Then $\L_1=\p^{2}-2x^{-2}$, this
is the singular part of the Treves test for the KdV hierarchy, see [1].

Let $n=3$ and $N=1$. Then $L_1=\p^3-3x^{-2}\p+3x^{-3}$ which gives the
singular parts of $u_1$ and $u_0$ for the test (i) (Eq. (2)).

Let $n=3$ and $N=2$. Then $L_1=\p^3-6x^{-2}\p+12x^{-3}$ which
corresponds to the test (ii) (Eq. (3)).

What about the second, conjugated, group of tests? As we know, it is generated
by the operators $(-1)^n\tL_N^*$. The singular parts are
$$(-1)^n\tL_N^*=\(\p+{1\over x}\)^{-N}\p^n\(\p+{1\over x}\)^{N}$$$$
=\(\p+{1\over x}\)^{n-1-N}\(\p-{n-1\over x}\)\(\p+{1\over x}\)^{N},
~0\leq N<n.\eqno{(7)}$$ Notice that both operators, (6) and (7),
coincide when $N=1$: $$(\p-x^{-1})\p^n(\p-x^{-1})^{-1}=
(\p+x^{-1})^{-1}\p^n(\p+x^{-1})$$ since $(\p+x^{-1})(\p-x^{-1})=\p^2$.
Therefore, one of tests of the first group ($N=1$) coincides with
the corresponding conjugated test. For the KdV (n=2) one has only
one test, for the Boussinesq ($n=3$) one has 3 different tests, and
so on.\\

{\bf 3.} In a recent work by Morosi and Pizzocchero [3], another method
of the proof of the necessity of the Treves criterion is suggested (they
considered the case of KdV, $n=2$). It
is based on the multiplicative representation of the operator $L$, as a
product of first-order differential operators, rather than in the dressing
form we exploited. They notice that if the term $u$ in the
operator is replaced by the Laurent series $\tu(x)$ satisfying the
Treves condition, then the operator is B\"acklund equivalent to an
operator without singularities. Then, they use the fact that for two
B\"acklund equivalent operators $L_1$ and $L_2$ the first integrals
$\res_\p L_1^{m/n}$ and $\res_\p L_2^{m/n}$ differ by a derivative of
a differential polynomial (see below) and, therefore, $$\res_x\res_\p \tL_1^
{m/n}=\res_x\res_\p \tL_2^{m/n}$$ and the fact that if an operator has no
singulariries then this quantity is zero. They conclude that for the given
operator this is zero, too. The relation between two method, at least
for singular parts of operators is clearly seen on the same formula (6).
There are two representations of the operators $L_N$ there: the first is
the dressing formula, the second is the product of first-order
operators. Two such operators with distinct $N$ differ by a cyclic
permutation of the first-order operators, i.e., they are B\"acklund
equivalent. One of them (when $N=0$) has no singularities. For the
conjugated operators, we use the formula (7).

{\it Example.} Let $n=3$. We have
$$-\tL_2^*=(\p-2x^{-1})(\p+x^{-1})(\p+x^{-1})=\p^3-6x^{-1}\p$$ which is
the singular part of the test (ii$^*$). How can we find the whole series
(4), following the logic of [4], after we got the singular part?
Firstly, we can write
$$-L_0^*=(\p+x^{-1}+v_1(x))(\p+x^{-1}+v_2(x))(\p-2x^{-1}-v_1(x)-v_2(x))$$
where $v_1(x)=\sum_0^\infty v_{1i}x^i/i!$ and
$v_2(x)=\sum_0^\infty v_{2i}x^i/ i!$. Then we must require that
this operator be regular. After some rather lengthy but
straightforward computations, one can show that this requirement
is equivalent to $v_1(x)$ and $v_2(x)$ being $O(x^2)$, i.e., the
series should start with terms with $x^2$. After that we calculate
$$-L_2^*=(\p-2x^{-1}-v_1(x)-v_2(x))(\p+x^{-1}+v_1(x))(\p+x^{-1}+v_2(x))$$$$
=\p^3+\tu_1^*(x)\p+\tu_0^*(x).$$ We can find that the only restriction
which should be imposed on the series for $\tu_1^*(x)$ and $\tu_2^*(x)$
in order they can be expressed in terms of $v_1(x)$ and $v_2(x)$ are
$u_{01}^*=0$ and $u_{02}^*=u_{13}^*/3$.

As it was said earlier, this guarantees that
$$\res_x\res_\p (L_2^*)^{m/3}=\res_x\res_\p (L_0^*)^{m/3}=0$$ if the
lemma is proven:\\

{\bf Lemma.} If two $n$th order operators are B\"acklund equivalent,
$$L_1=(\p+w_1)...(\p+w_n),~L_2=(\p+w_2)...(\p+w_n)(\p+w_1)$$ where
$w_i$ belong to a differential algebra, then $\res_\p L_1^{m/n}$ and
$\res_\p L_2^{m/n}$ differ by a derivative of an element of the same
differential algebra.\\

{\it Proof.} Evidently, $L_1(p+w_1)=(\p+w_1)L_2$, or $L_2=(\p+w_1)^{-1}
L_1(\p+w_1)$. This easily implies $$L_2^{m/n}=(\p+w_1)^{-1}L_1^{m/n}
(\p+w_1)=L_1^{m/n}(\p+w_1)(\p+w_1)^{-1}$$$$+[(\p+w_1)^{-1},L_1^{m/n}
(\p+w_1)]=L_1^{m/n}+[(\p+w_1)^{-1},L_1^{m/n}(\p+w_1)].$$ It remains
to take the $\res_\p$ of this equality and to notice that the residue
of the commutator of two operators with coefficients in some
differential algebra is a derivative of an element of the same algebra.
$\Box$

In our case, the differential algebra is the algebra of the Laurent
series, and we obtain the test (ii$^*$), Eq. (4).
\\

{\bf References.}\\

\noindent {\bf 1.} Dickey, L. A.:  On a generalization of the Treves
criterion for KdV hierarchy to higher  hierarchies, {\it Lett.in Math.Phys.}
{\bf 65} (2003), 187-197.\\
{\bf 2.} Treves, F. An algebraic characterization of the Korteweg- de Vries
hierarchy, {\it Duke Math.J.} {\bf 108}(2) (2001), 251-294.\\
{\bf 3.} Morosi, P. and Pizzocchero, L.: On a theorem by Tr\`eves,
arXiv: nlin.SI/0405007 (2004), 1-7.\\

\end{document}